\documentclass[]{iopart}
\usepackage{amssymb}
\usepackage{graphics}
\newcommand{\di}{\mathrm{d}}
\newcommand{\mrm}[1]{\mathrm{#1}}
\newcommand{\az}{\mbox{\boldmath$\alpha$}_{az}}
\begin{document}
\title{Subthreshold antiproton production in proton-carbon reactions}

\author{V I Komarov$^1$, H M\"uller$^2$  and A Sibirtsev$^3$}
\address{$^1$ Joint Institute for Nuclear Research, LNP, 141980 Dubna,
	Russia}
\address{$^2$ Institut f\"ur Kern- und Hadronenphysik,
        Forschungszentrum Rossendorf, D-01314 Dresden, Germany}
\address{$^3$ Institut f\"ur Kernphysik, Forschungszentrum J\"ulich,
        D-52425 J\"ulich, Germany}
\ead{H.Mueller@fz-rossendorf.de}

\begin{abstract} Data from KEK on subthreshold $\bar{\mrm{p}}$ as well
  as   on  $\pi^\pm$  and  $\mrm{K}^\pm$~production in  proton-nucleus
  reactions   are described at   projectile  energies between 3.5  and
  12.0~GeV. We use  a model which  considers a hadron-nucleus reaction
  as  an incoherent  sum over   collisions of  the projectile with   a
  varying number of target  nucleons.  It samples complete events  and
  allows thus for   the simultaneous  consideration of   all  particle
  species measured.  The overall   reproduction of the data   is quite
  satisfactory.   It   is  shown  that   the   contributions from  the
  interaction of the projectile with groups of several target nucleons
  are decisive for the description  of subthreshold production.  Since
  the collective features of subthreshold production become especially
  significant far below   the threshold, the  results are extrapolated
  down to COSY energies. It is concluded that an $\bar{\mrm{p}}$~measurement
  at ANKE-COSY  should be  feasible,  if the  high background of other
  particles can be efficiently suppressed.
\end{abstract}
%
\section{Introduction\label{intro}}

Subthreshold particle production in a  nuclear reaction is  understood
as production below the energy threshold  of the considered process in
a   free nucleon-nucleon  (NN) collision.   It   is  thus a  nuclear
phenomenon which may  be explained by  rather different assumptions on
the properties of nuclear matter and on the interaction dynamics.
Hereby it  is an open question   to what extent  subthreshold particle
production   is governed by  properties  of  the nuclear ground  state
wavefunction and to what extent by the dynamical properties of nuclear
matter, not reflected in the ground state description.  The problem is
far from   a final  solution at    present and  evidently   requires a
systematic  study  of high-momentum  transfer  processes,  among  them
subthreshold particle production.

At COSY  a research programme is in  progress which is devoted  to the
systematic     investigation of   subthreshold   particle  production.
Subthreshold production of $\mrm{K}^+$~mesons \cite{cosy18} was one of
the    foundations  for  the   design  and    building   of   the ANKE
spectrometer~\cite{barsov01},  which   allows for   the measurement of
$\mrm{K}^+$~meson production cross sections under  the condition of an
extremely  high  background  of   other  particles.  Recently,   first
results~\cite{koptev01,buescher02a,buescher04}   on  $\mrm{K}^+$~meson
production  have       been    published.   Moreover,       it     was
proposed~\cite{cosy21} to  extend  the    research programme  to   the
subthreshold   production    of   $\mrm{K}^-$~mesons.   This   made  a
considerable extension  of  the  ANKE~spectrometer necessary,  because
additional equipment for the registration of negative particles had to
be installed.

Following  this line, in this  paper  the question is  examined if the
investigation of subthreshold $\bar{\mrm{p}}$~production is manageable
at COSY energies.  The  highest energy of $2.8\,\mrm{GeV}$ reached  at
COSY up to now is far  below the threshold energy of $5.63\,\mrm{GeV}$
for  $\bar{\mrm{p}}$  production in NN~interactions.    This makes the
measurement a real challenge due to  the low cross  section and due to
the high background of  other particles.  From  the physical point  of
view,  however, deep subthreshold production  far  below the threshold
becomes especially interesting, because the  collective aspects of the
phenomenon can   be  expected  to become  more   and  more dominating.
Indeed,  the created  mass    in  case  of  $\bar{\mrm{p}}$~production
($1.876\,\mrm{GeV}$)      is    essentially    higher    compared   to
$\mrm{K}^+$~production    ($0.67\,\mrm{GeV}$),    and  the   energy is
$2.83\,\mrm{GeV}$        below         the       NN~threshold  if    a
$\bar{\mrm{p}}$~production  experiment   would   be   carried out   at
$2.8\,\mrm{GeV}$.  Such a  difference is significantly larger than the
value of   $0.58\,\mrm{GeV}$ achieved     so far   at   ANKE   in  the
$\mrm{K}^+$~production    experiments       at       an   energy    of
$1\,\mrm{GeV}$~\cite{koptev01}.

First   measurements   of subthreshold   $\bar{\mrm{p}}$~production in
proton-nucleus collisions at higher energies had been carried out
\cite{chamberlain55,chamberlain56,elioff62,dorfan65a} a  few   decades
ago.  Then investigations in nucleus-nucleus collisions
\cite{baldin88,caroll89,shor89,schroeter93,schroeter94}   and     more    
recent         studies       of         proton-nucleus       reactions
\cite{lepikhin87,chiba93,sugaya98} followed.  Most of the descriptions
proposed so far are based on transport calculations
\cite{li94a,batko91,huang92,cassing92,teis93,teis94,cassing94a,batko94,hernandez95,sibirtsev98},
thermodynamical considerations  \cite{koch89,ko88,ko89,dyachenko00} or
multi-particle interactions \cite{danielewicz90}. All these approaches
consider     the  $\bar{\mrm{p}}$~spectra   without   any  relation to
measurements of the  other reaction channels.  At KEK \cite{sugaya98},
however,  the  spectra of  $\pi^\pm$   and  $\mrm{K}^\pm$~mesons  were
measured together with those of the antiprotons.  The present approach
is distinguished   by considering simultaneously  all  these  reaction
channels.  This allows for  a more comprehensive determination  of the
model parameters and makes a prediction for COSY more profound.

It  should be  stressed that     the investigation of     subthreshold
production  in connection with the other  reaction channels is also of
great importance for the understanding of this process.  Fragmentation
of the target  residue is such  a distinguished channel having a large
cross section.  In  a participant-spectator picture  there seems to be
rather a weak connection  between the interaction of  the participants
leading  to  particle production and the   excitation  of the residual
nucleus during  this process.  At  subthreshold energies, however, the
competition between these  two energy consuming  processes may heavily
influence   the  cross sections observed.   In  \cite{muellerh92} this
question was considered  for proton-induced subthreshold production of
$\mrm{K}^+$ mesons.  There,  the interplay of subthreshold  production
and the fragmentation of the  residual nucleus was investigated in the
Rossendorf  collision (ROC)  model,   which allows  the   treatment of
hadronic   and nuclear reactions in  a  unified way.   Here, we extend
these considerations   to  the case   of  subthreshold $\bar{\mrm{p}}$
production.  Both  the fragmentation of the  residual  nucleus and the
interaction  of the nucleons  participating  in the scattering process
are  treated  on  the   basis of   analogous assumptions.  Even   more
important, the phase-space of  the complete final state  including the
reaction products of the projectile-participant interaction as well as
the fragments   of the   decay  of  the  spectator  system is  exactly
calculated.  To  the best of  our knowledge  this feature seems  to be
unique to the ROC model.

In   the ROC model the  nuclear   residue becomes  excited during  the
reaction due  to  the  distortion of  the  nuclear   structure by  the
separation of  the participants  from the  spectators and  due to  the
passage of the reaction products through the spectator system. In this
global  way final-state interactions   are taken into  account without
making special  assumptions  concerning  re-absorption, re-scattering,
self-energies,  potentials etc.  for the various  particle types.  The
discussion of   these properties  is   rather controversial.   If  one
considers  {\em  e.g.}    re-absorption of   $\bar{\mrm{p}}$  then the
treatment  reaches        from    no    explicit         consideration
\cite{danielewicz90,shor90} over  a   global  factor  0.1 for   p+Cu
\cite{batko91}   to taking the     mean    free path  into     account
\cite{huang92}.  For  Si+Si  the survival  rates  of antiprotons are
according to \cite{teis93,teis94,li94a}  in the region of  several~\%,
while in \cite{kahana93} a reduction of about  50\% for p+Au and Si+Au
from an  analysis  of  $\bar{\mrm{p}}$~production  at  higher energies
\cite{e802-ab93} has been reported.  The authors \cite{sugaya98}
conclude from an analysis of  the target-mass dependence of their data
that re-absorption of  $\bar{\mrm{p}}$  is much smaller  than expected
from  the corresponding $\bar{\mrm{p}}$N cross   section.  Thus, it is
far from  being clear to  what degree antiprotons  are absorbed in the
nuclear medium.

The plan of the paper is as  follows.  In section~\ref{model} the main
ingredients of  the  ROC model  are explained, which   is used for the
calculations   to be     presented.  Section~\ref{comp}  contains    a
comparison of   theoretical  and  experimental   results for  particle
production  and fragmentation  with  special emphasis  on subthreshold
$\bar{\mrm{p}}$  production.   A prediction  of  the antiproton  cross
section at COSY energies and a  comparison with the cross sections for
$\pi^-$  and $\mrm{K}^-$ production is  made in section~\ref{COSY}.  A
summary is given in section~\ref{sum}.

\section{The model\label{model}}
The ROC model is implemented as a  Monte Carlo generator which samples
complete events for hadronic as well as nuclear reactions. It makes no
detailed  assumptions  on   the   intra-nuclear  development  of   the
interaction process, but calculates instead the statistical weights of
the possible final states. The dynamics of the  reaction is taken into
account in form of empirical  functions which modify the population of
the final states.  This is in contrary  to transport models, where the
interaction of the projectile  with target nucleons and the subsequent
interactions  of  particles originating  from primary  collisions with
further target nucleons are  modelled.  The ROC approach does not need
any parameterisation  of elementary cross sections.  Its applicability
is not restricted at higher energies  by the growing number of unknown
elementary cross sections.  All dynamic information  is gathered  in a
few  parameters which  are either constant    or change smoothly  with
energy  and/or target mass.     The energy necessary  for subthreshold
production stems from interactions  of the projectile with few-nucleon
groups, also called  clusters in  the following,  and  from the  Fermi
motion of these clusters.

The  model was  successfully tested   for pp  interactions  up to  ISR
energies in \cite{muellerh95,muellerh01}, while nuclear reactions were
considered               in                     the             papers
\cite{muellerh92,muellerh91,muellerh95a,muellerh96}. In the  following
the basic  ideas of  the    ROC model  relevant for  this paper    are
summarised.

The cross section  of  the interaction of  an  incoming hadron  with a
target nucleus $(A,Z)$ consisting of $A-Z$ neutrons and $Z$ protons is
calculated  as an incoherent  sum over  contributions  from a  varying
number  $a$  of nucleons (thereof   $z$  protons) participating in the
interaction
\begin{equation} \label{dsigma}
  \di\sigma(s)=\sum_{a=1}^{A}   \sum_{z=max(0,a-Z)}^{min(a,Z)}
  \sigma_{az} \frac{\di W(s;\az)}{\sum_{\az}\int
    \di W(s;\az)}\,.
\end{equation}
Here, $s=P^2$ denotes  the square of the  centre-of-mass energy of the
projectile-target  system  with   $P=(E,\vec{P})$   being  the   total
four-momentum, $\sigma_{az}$  stands for the  partial cross section of
the interaction  of  the projectile with  a cluster   $(a,z)$, and the
quantities $\di W(s;\az)$  describe the relative probabilities  of the
various final channels $\az$.

The partial  cross    sections $\sigma_{az}$ account   for  sequential
collisions between projectile and $a$ nucleons  of the target.  Such a
sequence  is   treated in  the ROC model  as   a collision between the
incident particle and a cluster consisting of  $a$ nucleons capable of
sharing their energy in close analogy with the virtual clusters of the
cooperative model
\cite{knoll79,knoll80,bohrmann81,shyam84,shyam86,knoll88,ghosh90,ghosh92}.
The $\sigma_{az}$ are calculated using a modified version of the Monte
Carlo   code~\cite{shmakov88}  which  is  based   on   a probabilistic
interpretation of  the  Glauber theory~\cite{glauber70}.  We  use  the
profile function
\[
\Gamma_A (b)=\int \left[1-\prod_{i=1}^A (1-p_i)\right]
        \prod_{i=1}^A \rho_A(\vec{r}_i) \di^3r_i
\]
of the  considered nucleus,  which  depends   on the  nucleon  density
$\rho_A(\vec{r}_i)$ and the probability
\[
p_i=\exp(-d_i^2 \pi/\sigma_{NN})
\]
for an interaction of the projectile and the $i$th target nucleon with
$d_i$  being the   distance between the   interacting  particles.  The
nucleon density \cite{elton61}
\begin{equation} \label{rho}
  \rho_A(\vec{r})                                              \propto
  (1+\eta[1.5(f^2-e^2)/f^2+e^2r^2/f^4])\exp(-r^2/f^2)
\end{equation}
of light  nuclei $A <  20$ can be  derived from a standard shell model
wavefunction         with  $\eta=(A-4)/6$   ,    $f^2=e^2(1-1/A)$  and
$f=1.55\,\mrm{fm}$.    Then  the   NN~cross  section  $\sigma_{NN}$ is
adapted such that the integral of the profile function over the impact
parameter $b$ reproduces the total inelastic $pA$ cross sections
\[
\sigma_{pA}^\mrm{in} = \int \di^2b \Gamma_A(b),
\]
which  are approximately    constant   in the  energy   region   under
consideration.    The same calculation  yields  also the partial cross
sections $\sigma_{az}$ we are interested in (for further details see
\cite{shmakov88}).

The  relative  probabilities of  the various final   channels $\az$ in
(\ref{dsigma})
\begin{equation} \label{dWaz}
  \di W(s;\az) \propto \di L_{n} (s;\az) \rho_A(\vec{P}_R) T^2(\az)
\end{equation}
are given by the Lorentz-invariant phase-space factor $\di L_n(s;\az)$
multiplied  by  the square  of  the empirical reaction  matrix element
$T^2(\az)$ responsible for the collision dynamics. The Fermi motion is
implemented via the   momentum distribution  of the residual   nucleus
$\rho_A(\vec{P}_R)$,  which  is made a   function   of the number   of
participants $a$.  It is taken as a Gaussian having a width of
\begin{equation} \label{sigA}
   \sigma_a=\sqrt{a(A-a)/5/(A-1)}\,\,p_F
\end{equation}
in accordance with  the  independent particle model~\cite{goldhaber74}
with $p_F$  being the Fermi-limit   of the  nucleus  considered.    No
special                                                  high-momentum
component~\cite{shor90,geaga80,ciofi96,benhar89,sick94,sibirtsev97} is
used in this paper.

The Lorentz-invariant phase-space is  defined as the integral over the
four-momenta  of  the  $n$  primarily produced  final  particles  with
energy-momentum conservation taken into account
\begin{equation} \label{phase-space1}
  \di  L_n(s;\az)   =  \prod_{i=1}^n \di^4 p_i \,
     \delta (p_{i}^{2}-m_{i}^{2}) \,  \delta^4 (P-\sum_{i=1}^n p_i).
\end{equation}
Here,   the  four-momentum of   the    $i$-th particle  is denoted  by
$p_i=(e_i,\vec{p}_i)$ with $p^2_i=m^2_i$.

For       numerical   calculations     the     $\delta$-function    in
equation~(\ref{phase-space1}) has  to be removed  by introducing a new
set of $3n-4$ variables to replace the $3n$ three-momentum components.
It is  reasonable to choose  a  set of  variables,  which reflects the
underlying   physical picture   of   the interaction  process.   Using
recursion  \cite{byckling73} equation~(\ref{dWaz}) can be rewritten in
the form (see Appendix)
\begin{equation} \label{dWaz1}
   \di  W(s;\az)  \propto \frac{\di^3P_R}{2E_R} \rho_A(\vec{P}_R)
   \di W_R(M_R^2)\di W_P(M_P^2).
\end{equation}
Here, the four-momenta  of the  nuclear residue  $P_R=(E_R,\vec{P}_R)$
and of the participants
\begin{equation} \label{dPP}
P_P=P-P_R=(E_P,\vec{P}_P),
\end{equation}
yield the corresponding   invariant masses according  to $M_R^2=P_R^2$
and $M_P^2=P_P^2$.  The integral over the Fermi motion $\di^3P_R
\rho_A(\vec{P}_R)   /2E_R$     separates 
the phase-space of the $n_R$ nuclear fragments
\begin{equation} \label{dWR}
    \di W_R(M_R^2) = \di   M_R^2  \di L_{n_R}(M_R^2) T_R^2(\az^R)
\end{equation}
from  the   phase-space  
\begin{equation} \label{dWP}
   \di W_P(M_P^2) = \di L_{n_P}(M_P^2) T_P^2(\az^P)
\end{equation}
of  the $n_P=n-n_R$ final    particles  arising from the   participant
system. In (\ref{dWR})  and (\ref{dWP}) the matrix  element
\begin{equation} \label{ME}
  T^2(\az)=T_R^2(\az^R)   T_P^2(\az^P)
\end{equation}
is  split  into  two  factors describing  residue  fragmentation  into
channel  $\az^R$  and  participant  interaction resulting  in  channel
$\az^P$,  accordingly.  There is,  however,  a  strong kinematic  link
between participants and spectators, since invariant mass $M_P$ of the
participant  system,  invariant mass  $M_R$  of  the  residue and  the
relative  kinetic  energy  $\sqrt{s}-M_P-M_R$  of these  two  particle
groups  are connected by  energy-momentum conservation.   For particle
production  to proceed  the invariant  mass of  the  participants must
exceed  the corresponding  threshold  value $M_P^{th}$  which in  turn
depends  on the  number of  participants.  The  heavier  the effective
target  is  the more  energy  is  available  for particle  production.
Another way  to reach the threshold  value goes via  the Fermi motion,
since the participant mass is a function of the momentum vector of the
nuclear residue.   The excitation energy  of the target  residue comes
usually  into   play  via  the  spectral  function   (see  {\em  e.g.}
\cite{ciofi96}) derived  from electron scattering data.  Unique to the
ROC model, however, is the treatment of  the spectator system in close
analogy with the participant  subsystem.  The ROC model calculates the
complete  final state of  both the  participants  and the  spectators.
Thus, the huge amount of final channels of the spectator fragmentation
influences directly the final state of the participant system and vice
versa.

The term $\di W_P(M_P^2)$  in (\ref{dWaz1}), (\ref{dWP}) describes the
interaction of the incoming   proton with the group  of  participating
target   nucleons.  Such a projectile-cluster   reaction is treated in
complete analogy to a hadronic reaction.  In a first step intermediate
particle groups called fireballs (FBs)  are produced, which decay into
so-called  primary   particles.   The  primary   particles define  the
channels for which the weights (\ref{dWP}) are calculated.  Among them
are resonances, which decay subsequently into  stable hadrons. In case
of the presence  of several nucleons among the  decay products of a FB
they   may  form nuclear fragments    with  equal probability  for all
possible channels.

The  dynamical input of  the reaction is  implemented by the empirical
transition matrix element

\begin{equation} \label{A2P}
  T^2_P(\az^P)=  T^2_{\mrm i}\,   T^2_{\mrm {qs}}\, T^2_{\mrm  {ex}}\,
  T^2_{\mrm t}\, T^2_{\mrm l}\, T^2_{\mrm {st}}\,,
\end{equation}
which describes  the interaction process  $T^2_{\mrm  i}$ resulting in
the production of  a   varying number $N$   of FBs  ($N \ge  2$),  the
production   of    hadrons  $T^2_{\mrm {qs}}$ via     the  creation of
quark-antiquark ($q\bar{q}$) pairs, the invariant-mass distribution of
the  FBs    $T^2_{\mrm  ex}$,   the  transverse  $T^2_{\mrm    t}$ and
longitudinal  $T^2_{\mrm l}$  momentum distribution  of  the FBs, and,
finally, some factors $T^2_{\mrm {st}}$  necessary for the calculation
of the statistical weights.  The interaction is assumed to proceed via
colour exchange leading to the removal  of valence quarks or of gluons
from  the interacting hadrons.  Additional  up, down and strange quark
pairs are created in the ratio
\begin{equation} \label{lamba}
u:d:s = 1:1:\lambda_s
\end{equation}
with $\lambda_s=0.15$.  They form  the  varying number  of  FBs, which
subsequently decay into the final hadrons.   The transverse momenta of
the  FBs  are  restricted  by   an  exponential cut-off  (longitudinal
phase-space) according to
\begin{equation} \label{At}
  T^2_{\mrm t} = \prod_{I=1}^N \exp(-\gamma P_{\mrm {t,I}})
\end{equation}
with the mean $\bar{P}_{\mrm t}=2/\gamma$.
Two leading FBs, the  remnants  of the incoming hadron
and of the cluster, get in  the mean larger longitudinal momenta than
the   central FBs by weighting the events with
\begin{equation} \label{Al}
  T^2_{\mrm l} = (X_1 X_2)^\beta.
\end{equation}
Here, the light-cone variables $X_1=(E_1+P_{\mrm    z,1})/(e_b+p_{\mrm
z,b})$ and  $X_2=(E_2-P_{\mrm z,2})/(E_P-P_{\mrm z,P})$ are  used with
the four-momentum of the projectile given by $p_b=(e_b,\vec{p}_b)$ and
that of the participants by  (\ref{dPP}).  Each FB is characterised by
two  parameters, a temperature $\Theta_{FB}$   and a volume  $V_{FB}$.
The temperature    determines  the  relative kinetic  energy    of the
particles the FB decays into via
\begin{equation} \label{AexP}
  T^2_{\mrm ex}(\Theta_{FB}) =
        \prod_{I=1}^N (M_I/\Theta_{FB})  K_1(M_I/\Theta_{FB}),
\end{equation}
while the volume defines the interaction  region and influences mainly
the particle multiplicity via the  statistical factor $T^2_{\mrm {st}}
\propto V_{FB}^{n_P-1}$. In (\ref{AexP}) $K_1$ stands for the modified 
Bessel function. Final hadrons are built-up by random recombination of
the available  quarks during  the  decay of  the FBs.   This procedure
ensures   automatically  the   conservation  of   all internal quantum
numbers.  Resonances decay later on  until the final state  consisting
of stable particles is reached. For a  more detailed discussion of the
hadronic  matrix element the  reader is referred to \cite{muellerh01}.
In  this    paper  we use    the   same  set   of   parameters  as  in
\cite{muellerh01}   for  the description of    the  interaction of the
projectile with  a single target nucleon.   Most of the other terms in
(\ref{dsigma}), however, contain clusters consisting of several target
nucleons.  The   basic parameters of the  FBs   emerging from  such an
interaction are fixed by scaling the volume with the number of cluster
nucleons $a$ and  the temperature parameter with $a^{-1/3}$  according
to
\begin{equation} \label{sca}
V_{FB} = V_{FB}^0 \,a \quad \mbox{and} \quad
\Theta_{FB} = \Theta_{FB}^{max}\,a^{-1/3}.
\end{equation}

It remains to  consider the target residue, the  structure of which is
strongly  disturbed  by   the  projectile-participant   reaction   and
subsequent final-state interactions.  This leads to the excitation and
decay  of the  spectator  system, which  is  characterised  by the two
parameters temperature $\Theta_R$ and volume $V_R$  in the same way as
the FBs  emerging from the   projectile-participant interaction.   The
part of the matrix element responsible for the residue fragmentation
\[
   T_R^2(\az^R) = T^2_{\mrm {ex}}(\Theta_R)\, T^2_{\mrm {st}}(\az^R)
\]
is identical with the corresponding factors (\ref{A2P}) applied to the
hadronic FBs.  In order  to restrict the excitation energy transferred
to the residue we use the asymptotic approximation of (\ref{AexP}) for
large mass $M_R$ and small temperature $\Theta_R$
\begin{equation} \label{Aex}
  T^2_{\mrm ex}(\Theta_R) = \sqrt{M_R/\Theta_R} \,  \exp(-M_R/\Theta_R).
\end{equation}
An impact  parameter  dependence is    assumed  for the    temperature
parameter.   This seems    to be  reasonable,  because   a  peripheral
collision with only  few   participating nucleons should excite    the
nuclear  residue  much less  than   a   central collision  with   many
participants.  As a first guess we use
\begin{equation} \label{ThetaR}
   \Theta_R = \Theta_R^{max}\, [1-\exp{(-a/\bar{a}A^{1/3})]}
\end{equation}
with  $\bar{a}$ as parameter,    here fixed  to  $\bar{a}=0.5$,  which
determines  how  fast   the maximal   temperature $\Theta_R^{max}$  is
reached with increasing number of participants $a$.

All factors still necessary for a  correct calculation of the relative
weights of the various channels are collected in complete analogy with
the corresponding factors for the FBs in the term
\begin{equation} \label{Ast}
  T^2_{\mrm{st}}(\az^R)=g(\az^R)
  \left(\frac{V_R}{\left(2\pi\right)^3}\right)^{n_R-1}
  \prod_{i=1}^{n_R}(2\sigma_i+1)\,2m_i.
\end{equation}
It  contains the spin  degeneracy factors  $(2\sigma_i+1)$, the volume
$V_R$  in which the particles  are produced with $V_R=4 \pi (A-a)
R_R^3/3$ determined  by   the radius  parameter $R_R$.    The quantity
$g(\az^R)$ is the degeneracy  factor for groups of identical particles
in the final state of the residue decay and prevents multiple counting
of identical states.

Temperature and volume determine as in the  case of hadronic reactions
the  number of final  particles and  their relative  energy.  The main
difference consists    in  the value  of   the  temperature parameter,
$\Theta_{FB}^{max}   \approx   300\,   \mrm{MeV}$   for hadronic   and
$\Theta_R^{max}  \approx  10\, \mrm{MeV}$  for nuclear   systems.  New
particles can  be produced  in  hadronic reactions, while  for nuclear
systems,  due  to the much   lower  temperature, the nucleons  of  the
initial state are recombined  into various fragments without producing
new hadrons.   The volume parameter defines  the distance  between the
fragments where the strong interaction   ceases to work.  This  volume
goes  into the calculation  of the weight of  the final state.  Due to
the long range of the Coulomb repulsion Coulomb energy is still stored
in the system  at those distances.   This energy is calculated  in the
Wigner-Seiz-approximation  (see  \cite{bondorf85}).      The    random
distribution of the Coulomb energy among  the charged fragments yields
the final momenta of the fragments at large distances.

In the  calculation  we assume  that  the fragments emerging  from the
residue  decay are stable against particle  decay.  The possibility of
hot fragments cooling down  by  subsequent particle emission is   also
implemented in the  model (see \cite{muellerh92}),  but it turned  out
that a  readjustment of  the volume  and/or  the temperature parameter
yields quite similar  results in both cases.  So  we decided in favour
of  the easier approach.  Moreover, for  a light nucleus as ${}^{12}$C
this question is of minor importance.

In concluding  this  section it should be   mentioned, that it  is the
calculation of  the  relative probabilities  $\di  W(s;\az)$, equation
(\ref{dWaz}), which   makes the difference   between the   cooperative
\cite{knoll79,knoll80,bohrmann81,shyam84,shyam86,knoll88,ghosh90,ghosh92}
and the ROC model, although the notion  of clusters is similar in both
approaches.  In  the cooperative  model the relative  probabilities of
the various channels are calculated  in the pure phase-space limit and
the   excitation  and  decay  of  the  target    residue is completely
neglected.     In contrast, the ROC model is    far  from being a pure
statistical approach, because it  contains specific information on the
properties  of nuclei  and  the  interaction of  the   projectile with
few-nucleon groups (clusters)  in  the  nucleus.  The model   includes
strong   assumptions   on the  amplitude  of   the  projectile-cluster
interaction  which lead  to  the restriction of  excitation energy and
momentum transfer to clusters compared to pure phase-space.  A further
nontrivial  assumption  concerns the   various  possibilities for  the
dissipation of  the transferred energy  inside the hit clusters either
as relative kinetic energy of the  decay products or as newly produced
particles.   In extreme   cases  the whole  energy can  be  completely
accumulated for the production of new particles.  The other extreme is
the  emission of  very   fast secondaries with  energies outside   the
kinematical limits of a projectile-(single)nucleon interaction without
producing new particles.    Also the supposition that  clusters behave
themselves like hadrons  with respect to  quark statistics is of great
importance for the  relative   weights of the various  final  reaction
channels.    All  these  assumptions  form  a   new  approach to   the
consideration of local excitations  and may, therefore, give essential
and definite  information    on  the  properties  of  nuclear   matter
(excitation probability,  distribution     of  deposited energy    and
transferred momenta, decay modes of few-nucleon fireballs etc).

\section{Comparison with data \label{comp}}
\begin{figure}
\begin{center}
   \resizebox{0.6\textwidth}{!}{%
   \includegraphics{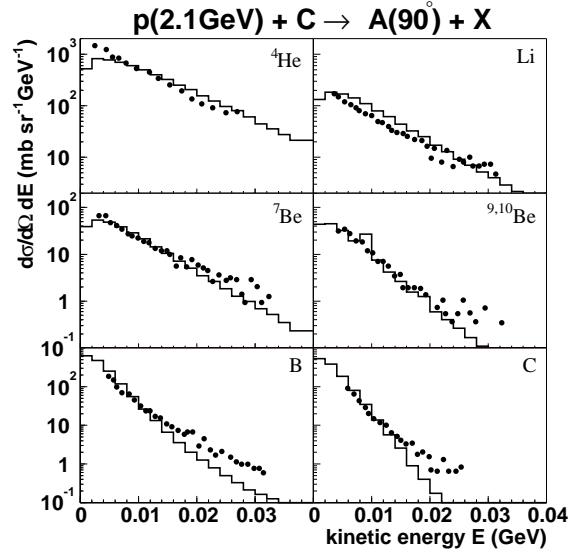}}
   \caption{\label{p21C} Differential cross section as function of the
   kinetic  energy  of fragments produced in  the  reaction of protons
   with carbon~\cite{westfall78}    (dots)   compared  to    ROC-model
   calculations (histograms).}
\end{center}
\end{figure}
\begin{figure}
   \resizebox{0.95\textwidth}{!}{
   \includegraphics*{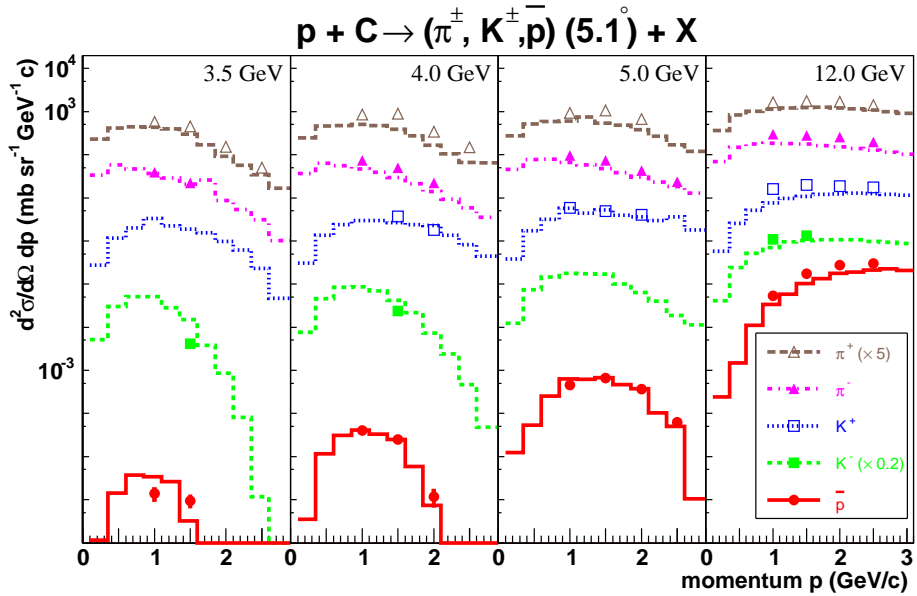}}
   \caption{\label{pCpBar}Momentum spectra of $\pi^\pm$, $\mrm{K}^\pm$
   and $\bar{\mrm{p}}$ \cite{sugaya98} (symbols) compared to ROC-model
   calculations  (histograms). Experimental and calculated results for
   $\pi^+$  and  $\mrm{K}^-$ mesons  are  multiplied  by  the  factors
   indicated in the legend.}
\end{figure}

In order to  fix the  parameters describing the  Fermi motion  and the
decay of the residue we consider  the energy spectra of fragments from
the  reaction   of    2.1~GeV   protons   with carbon    measured   at
$\mrm{90}^\circ$ (see figure~\ref{p21C}).  For   the volume  a  radius
$R_R=1.7\,\mrm{fm}$   is taken.  This  corresponds   to about $1/3$ of
normal nuclear density and is within  the region of break-up densities
used for   example  by  \cite{bondorf95}.  In   figure~\ref{p21C}  the
heaviest fragment, $C$, arises mainly  from the quasi-free interaction
of the   incoming proton with a single   target neutron  and  may gain
energy  only  from the Fermi  motion.   Thus, the   spectrum is highly
sensitive  to   the width  $\sigma_a$,  equation~(\ref{sigA}), of  the
Gaussian for the   Fermi  motion.  On  the other   hand, the  lightest
fragments  are decay products of the  residue and their kinetic energy
in the  laboratory  is a superposition  of  Fermi motion  and relative
kinetic energy of the residue fragments. A satisfactory description of
the  fragment   spectra   in  figure~\ref{p21C}   is   achieved  using
$\Theta_R^{max}=12\,\mrm{MeV}$,      equation~(\ref{ThetaR}),      and
$p_F=320\,\mrm{MeV/c}$, equation~(\ref{sigA}).  No attempt was made to
improve the description of the  spectra by implementing an  additional
high-momentum  component,  although especially the high-momentum parts
of  the  spectra  of the  heavier   fragments are underestimated.  The
spectator-participant  picture is  an  idealisation, and  not only the
nuclear  wavefunction  but     also  secondary interactions    between
spectators  and participants  might   be responsible for the  observed
deviations.   All the following calculations are  carried out with the
above  fixed   values for $\Theta_R^{max}$  and   $p_F$.  It should be
stressed  in this connection that   temperature $\Theta_R$ and  volume
parameter $R_R$  of the nuclear  residue are highly correlated, a fact
already   observed in  \cite{muellerh01} for   the   FB parameters  in
hadronic reactions.

In   figure~\ref{pCpBar}  the   momentum   spectra  of   $\pi^\pm$ and
$\mrm{K}^\pm$~mesons as well as of $\bar{\mrm{p}}$ are compared to the
KEK data~\cite{sugaya98}.  The overall agreement is quite satisfactory
in view of the large region of projectile  energies and the variety of
ejectile  species calculated with  one fixed  parameter set.  Particle
yields are   influenced  by the suppression  factor  $\lambda_s=0.15$,
equation~(\ref{lamba}), of   strange quarks and  by  the algorithm for
creating the final hadrons from the quarks produced in the first stage
of the    interaction  process.  Hadrons   are  built  up   in each FB
independently   according   to  the     rules   of   quark  statistics
\cite{anisovich73}  by randomly   selecting  sequences  of  $q$'s  and
$\bar{q}$'s.  A $q\bar{q}$ gives a meson, while baryons or antibaryons
are  formed  from $qqq$  or    $\bar{q}\bar{q}\bar{q}$.  From a  given
sequence of quarks the different  hadrons are formed according to  the
tables of the particle data group~\cite{PDG98}.  There is no parameter
which   directly determines   the   ratio  between  meson and   baryon
production as e.g. in the PYTHIA-LUND model
\cite{andersson83,andersson87,sjostrand87a,pythia94}.      Only     an
indirect influence   via the temperature and  the  volume parameter is
possible,  which change the relative  weights of the FBs in dependence
on their invariant mass and final particle multiplicity, respectively.

It is therefore quite remarkable that the general trend of the data is
well reproduced by using the parameters determined in
\cite{muellerh01} from  the consideration of  hadronic  interactions.
The moderate increase with energy of  the cross sections for the light
mesons, the steep increase of that for the  antiprotons as well as the
shift of  the maximum  in  the $\bar{\mrm{p}}$~spectra  towards higher
momenta is well reproduced by the calculations.
\begin{figure}
\begin{center}
   \resizebox{0.7\textwidth}{!}{
   \includegraphics{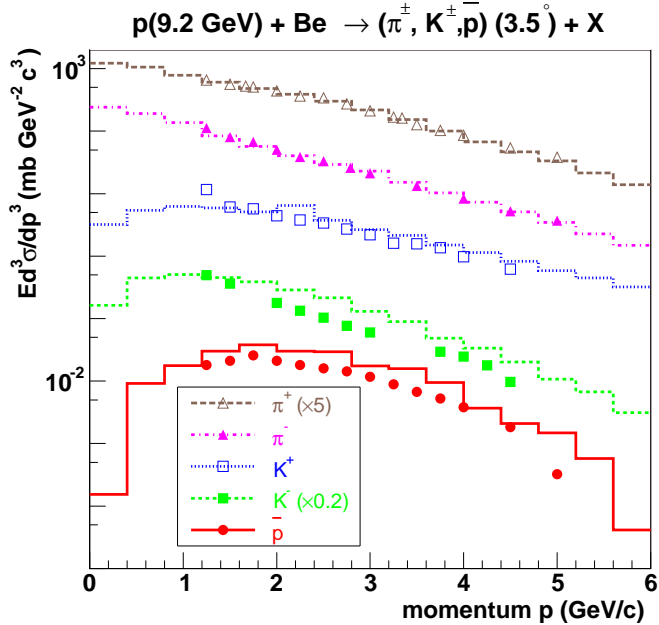}}
   \caption{\label{pBe}   Momentum spectra of $\pi^\pm$, $\mrm{K}^\pm$
   and  $\bar{\mrm{p}}$  \cite{vorontsov88}    (symbols)   compared to
   ROC-model calculations  (histograms).  Experimental and  calculated
   results  for $\pi^+$ and $\mrm{K}^-$  mesons  are multiplied by the
   factors indicated in the legend.}
\end{center}
\end{figure}

At  the highest  energies   the  ROC calculations obviously   tend  to
underestimate the data.  Therefore, we  compare in figure~\ref{pBe} as
a kind of cross-check a  similar data set from \cite{vorontsov88} with
ROC-model calculations.  Here, the  spectra are measured at a  similar
angle but  in a much wider  momentum region compared  to the KEK data.
In this  case the tendency  of  the ROC results goes  in the  opposite
direction and   overestimates   the  cross-sections for  the   heavier
ejectiles $\mrm{K}^\pm$ and $\bar{\mrm{p}}$.

The calculated cross sections are sensitive to the assumed temperature
of  the residual nucleus  (see  figure~\ref{energyDep}).  It should be
stressed that the temperature derived from the fragment spectra yields
also a  reasonable reproduction of the $\bar{\mrm{p}}$~cross sections.
This point is  of special  importance  since it demonstrates that  the
link between the realms of fragmentation and of hadron production made
by the ROC approach is obviously correct.

The  authors \cite{sugaya98}  interpret  their $\bar{\mrm{p}}$~data by
using the ``first-chance NN collision model'' from \cite{shor90} where
the internal  nucleon  momenta  were  extracted from  backward  proton
production \cite{geaga80}  as  a    superposition  of   two   Gaussian
distributions.    In   this   way   the  momentum  spectra    and  the
incident-energy  dependence    could  be  successfully   reproduced by
adapting  one  normalisation  parameter.    For  the similar   case of
subthreshold $\mrm{K}^+$~production       it  has    been,    however,
argued~\cite{sibirtsev95b}   that   the      contribution    from  the
high-momentum component should be negligible.

In the ROC model the ``first-chance  NN collision'' corresponds to the
term  in (\ref{dsigma}) describing  the  interaction of the projectile
with  a single  target   nucleon.  In table~\ref{table}  the  relative
contributions of the terms with a given  number of participants to the
total $\bar{\mrm{p}}$ production cross section  are summarised. It can
be seen  that the interaction  with a single  nucleon yields  the main
contribution only at the highest energy. The lower the energy the more
contribute the  terms with  several  target nucleons.   At  the lowest
energy,  $2.8\,\mrm{GeV}$, considered here  the  main contributions to
$\bar{\mrm{p}}$~production arise from interactions with three and four
target nucleons, while  quasi-free collisions are negligible.   At the
highest    energy, $12\,\mrm{GeV}$,   there   are  still  considerable
contributions from  proton-cluster reactions, although      quasi-free
collisions became the dominating process.  Thus, the interpretation of
the data by the ROC model is quite contradictory to the assumptions of
a ``first-chance  NN  collision model''.   From  the viewpoint of  the
ROC model  the key quantity   for understanding subthreshold  particle
production   is   the number  of    participating  nucleons.  A direct
experimental  determination     of this   number    for   subthreshold
$\bar{\mrm{p}}$~production is highly desirable as discussed in
\cite{muellerh96} for the case of $\mrm{K}^-$~production.

\begin{table}
   \caption{ROC  results for the  relative contributions  to the total
   $\bar{\mrm{p}}$ cross  sections  at  selected energies in   \% from
   terms containing various  numbers $a=1 \dots  6$ of participants as
   well  as for the total  cross sections $\sigma_{\bar{\mrm{p}}}$ and
   the  cross  sections  $\sigma_{\bar{\mrm{p}}}^{\mrm{nf}}$  with  no
   freezing of kinetic degrees of  freedom (see text).}  \label{table}
   \lineup \begin{indented}
   \item[] \begin{tabular}{@{}lllllllll}
   \br
   Energy(GeV) & 1 & 2 & 3 & 4 & 5 & 6 &
   $\sigma_{\bar{\mrm{p}}}(mb)$ & $\sigma_{\bar{\mrm{p}}}^{\mrm{nf}}(mb)$  \\
   \mr
   2.8& 0.0& 0.9&51.3&37.5&9.8&0.4 &7.8$\,10^{-10}$&6.3$\,10^{-11}$\\
   3.5&0.05&20.9&56.6&19.3&2.9&0.2 &1.5$\,10^{-6}$ &7.9$\,10^{-7}$ \\
   5.0&10.9&30.5&43.1&13.2&2.0&0.3 &2.9$\,10^{-4}$ &4.3$\,10^{-4}$ \\
  12.0&63.3&21.7&11.9& 2.6&0.4&0.05&0.10           &0.13           \\  
   \br
   \end{tabular}
   \end{indented}
\end{table}

A  certain similarity can  be, however, found between the ROC approach
and the multi-nucleon  mechanism  of \cite{hernandez95},  where  ``the
incoming  proton and  the interacting  nucleons  in the  target act as
sources of pions that merge to produce a nucleon-antiproton pair''. In
this approach the  number of participating  nucleons obviously plays a
similar role as in the ROC model.

Another interesting aspect of subthreshold  particle production is the
observation  that the formation of  light nuclei  in the final channel
alongside the  produced particle(s) has to  be taken into  account for
achieving a good reproduction  of  the measured  cross  sections.  The
formation of light nuclei leads   to the freezing of relative  kinetic
energy   of  the nucleons,  which   is then   available   for particle
production.  We are speaking here  about light fragments emerging from
clusters, not about the decay of the residual nucleus. This effect has
been discussed in  \cite{shyam84}  for  pion, in \cite{shyam86}    for
photon,   in        \cite{ghosh92}  for     $\mrm{K}^+$~production  in
nucleus-nucleus  reactions on the basis  of the cooperative model, and
in   \cite{muellerh96},  within   the  ROC model,    for  subthreshold
$\mrm{K}^-$~production  in proton-nucleus  interactions.  In order  to
demonstrate  the    importance of   this    effect for    subthreshold
$\bar{\mrm{p}}$ production we have in table~\ref{table} in addition to
the  ``normal''  ROC results  $\sigma_{\bar{\mrm{p}}}$ also the  cross
sections $\sigma_{\bar{\mrm{p}}}^{\mrm{nf}}$ with no freezing  listed.
At  the lowest energy considered the  difference  amounts to about one
order of  magnitude.  This is  relatively small  compared to the three
orders  of magnitude reported     by \cite{shyam84} for  the  reaction
${}^{12}\mrm{C}(85\mrm{MeV}/A) + {}^{12}\mrm{C} \to  \pi^+ +  X$.  The
energy of 2.8~GeV is still far above the total production threshold of
about 2.2~GeV for the coherent production at the whole target nucleus.
Contrary to the energy considered in \cite{shyam84} also the number of
competing channels  is   still   rather high.    This  diminishes  the
influence of  the   neglected  channels.   The  slight  increase    of
$\sigma_{\bar{\mrm{p}}}^{\mrm{nf}}$              compared           to
$\sigma_{\bar{\mrm{p}}}$ at the higher energies is due to the smoother
energy  dependence of the  relative probability  $\di W(s;\az)$ of the
considered  channel   and   the    smaller denominator   in   equation
(\ref{dsigma}) as   a result of neglecting  a  large number   of final
channels.

In  \cite{teis93,teis94,cassing94a,sibirtsev98} calculations have been
carried  out  in the framework of  transport  theory,  where questions
concerning baryon self-energies, elementary $\bar{\mrm{p}}$~production
amplitudes,             $\bar{\mrm{p}}$~potentials                 and
$\bar{\mrm{p}}$~re-absorption are of  importance for  the reproduction
of the experimental data.  In the  ROC model there is no equivalent of
these intra-nuclear properties implemented so far.  Nuclear properties
enter  the  ROC calculations only via  the  scaling (\ref{sca}) of the
temperature  and the volume of the  FBs and via  the excitation of the
target residue caused by secondary participant-spectator interactions.
Nevertheless, the spectra of all  particle species are comparably well
reproduced.  This result  should  be further explored  by  considering
heavier  target nuclei,  where  the problem of secondary  interactions
becomes  more important.   On   the   other hand, the     experimental
investigation of $\bar{\mrm{p}}$~production  on the lightest nuclei as
${}^3He$ and ${}^4He$ is of special interest, since the model predicts
in the  energy range  between  3 and 5~GeV just   the 3- and 4-nucleon
groups as the  main  source of $\bar{\mrm{p}}$~production.   Secondary
interactions of   the produced  particles and   the influence  of  the
excitation of the target residue are  practically absent in this case.
This makes the interpretation of such data more clear than in the case
of heavier target nuclei.

\section{Antiproton cross section at COSY energies \label{COSY}}
\begin{figure}
\begin{center}
   \resizebox{0.5\textwidth}{!}{
   \includegraphics{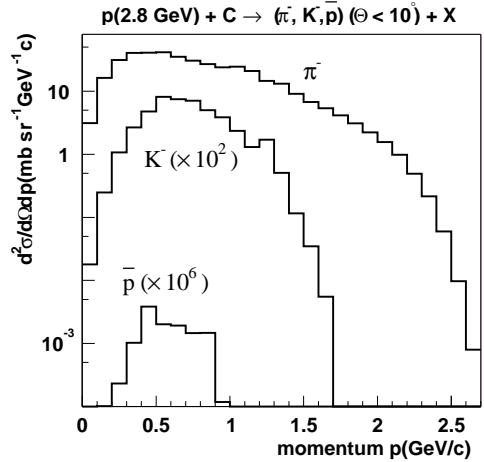}}
   \caption{\label{p28C}  Momentum spectra   of  negatively    charged
   hadrons calculated with the ROC model (histograms).}
\end{center}
\end{figure}
Figure~\ref{p28C} shows  a   ROC-model   estimate of  the   antiproton
spectrum   for a bombarding energy of   2.8~GeV  averaged over a polar
angle $\Theta <  10^\circ$.   In order  to emphasise  the experimental
difficulties of a measurement at such a  deep subthreshold energy, the
spectra of the other negative stable  particles are depicted, too.  In
the region of  the maximum in the  spectra around 500~MeV/c  about ten
orders of    magnitude more $\pi^-$~mesons  than   antiprotons must be
expected  what makes a possible  measurement  in addition to the small
$\bar{\mrm{p}}$ cross section a difficult task.

\begin{figure}
\begin{center}
   \resizebox{0.5\textwidth}{!}{
   \includegraphics{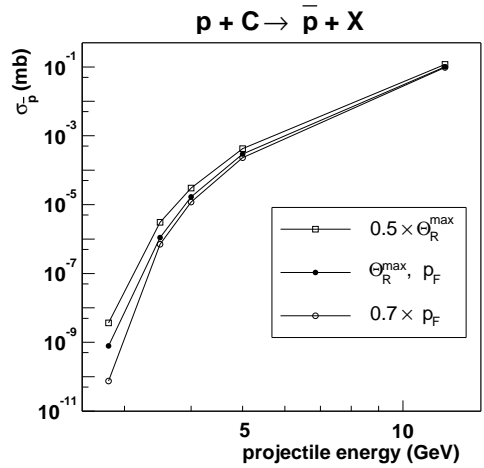}}
   \caption{\label{energyDep}         Cross        sections         of
   $\bar{\mrm{p}}$~production as  function  of incident energy between
   2.8 and 12.0~GeV.   The $\bar{\mrm{p}}$  cross sections  calculated
   for   three  parameter sets  (see text)  are  connected by straight
   lines.}
\end{center}
\end{figure}

In   figure~\ref{energyDep}   the energy    dependence  of  the  total
production cross    section    for $\bar{\mrm{p}}$   in     $\mrm{p} +
{}^{12}\mrm{C}$   interactions  is  plotted.   In the    energy region
considered the  cross section changes   by more than seven  orders  of
magnitude.  The importance of the  temperature of the residual nucleus
at low energies is demonstrated by  a calculation with the temperature
arbitrarily decreased  by a factor  of two.  This  increases the cross
section by about  half an  order of  magnitude at  the lowest  energy.
With increasing energy the influence  of this parameter diminishes and
becomes negligible above $10\,\mrm{GeV}$ projectile energy.  The Fermi
motion is  of   similar  importance for subthreshold   $\bar{\mrm{p}}$
production.  A decrease of $p_F$, equation~(\ref{sigA}), by about 30\%
makes  the  $\bar{\mrm{p}}$  cross  section   by  nearly one  order of
magnitude smaller.   Again  the influence  becomes  negligible at high
energies. These results underline once more how important it is to use
a model which allows the verification of  these aspects by independent
measurements as  has been done in  this paper.   It should be stressed
that the selected parameter range does  not reflect the uncertainty of
the considered parameters which  can be rather precise determined from
the   fragmentation   data~\cite{westfall78} (see  figure~\ref{p21C}).
Instead, it is the  aim to demonstrate the  sensitivity of the results
to the inclusion of the nuclear residue  into the considerations.  All
results in  the subthreshold region   from  approaches neglecting  the
excitation of   the nuclear residue  should be,  therefore, taken with
caution.

But  also the present  prediction   contains  of course   considerable
uncertainties. From our experience with existing experimental data, in
particular those shown in figures~\ref{pCpBar}  and \ref{pBe}, we know
that deviations between data and calculations  are usually less than a
factor  of two.  Since any drastic  change of the model feasibility is
not seen if the energy decreases from 3.5~GeV to 2.8~GeV we may expect
the same level of  uncertainty at this  lowest energy.  Certainly, for
the  present case of  a prediction rather far  below  the region where
experimental  data are available   it cannot  be  excluded that   some
processes or  factors not  included in  the ROC approach  may increase
their   significance with the  energy reduction.    A possible example
could    be the  coherent    production at the   whole  target nucleus
\cite{Alvaredo96}.  To be  on the sure side  we, therefore,  assume at
least  the order  of magnitude  of the  predicted cross section  to be
correct.

At the ANKE spectrometer an effective  detection of particles produced
at   angles $\lessapprox  10^\circ$ in    the  momentum range $0.2 -
1.0\;\mrm{GeV/c}$ is possible.  Assuming a luminosity of $10^{33}\,
\mrm{cm}^{-2} \mrm{s}^{-1}$  available  with a  carbon  strip  target 
and   a cross section  of  $1   \cdot  10^{-9}\,\mrm{mb}$ expected  at
$2.8\,\mrm{GeV}$  one gets  a   counting rate at   the  level  of five
hundreds  events per  week of   beam-time.   This estimate  shows  the
feasibility of studying $\bar{\mrm{p}}$~production  even at such a low
subthreshold energy as considered  here, if the experimental equipment
allows for an efficient suppression of the negative mesons.

\section{Summary \label{sum}}

Subthreshold particle production is  a collective phenomenon which  is
far from being  completely  understood. Data on  subthreshold particle
production can  be reproduced within the ROC  model by considering the
interaction  of the projectile  with  few-nucleon systems  in complete
analogy to the interaction with a single nucleon, also with respect to
high-momentum  transfer     processes.    It  is     the  simultaneous
consideration   of the  data for   all  particle types measured, which
distinguishes the  present  approach from   the  previous attempts  to
interpret  solely the  subthreshold $\bar{\mrm{p}}$~spectra from  KEK.
The  reproduction  of   the  data  requires no strong   distortion  of
antiprotons in nuclear matter, which might be expected due to the fact
that  the free  $\bar{\mrm{p}}N$  cross  section  is  larger  than the
corresponding ones for the   other particle species  considered.  This
finding     needs       further       confirmation     by    measuring
$\bar{\mrm{p}}$~spectra at  lower momenta than  the  region covered by
the  KEK  data where   the   $\bar{\mrm{p}}$N cross section  increases
further.

In   this     paper  the  feasibility   of     measuring  subthreshold
$\bar{\mrm{p}}$~production at COSY-ANKE is demonstrated.  It is argued
that new insight into  the mechanism of subthreshold  production could
be  gained    from an  experiment    nearly $3\,\mrm{GeV}$   below the
NN~threshold, where the   ROC-model  calculations  predict  increasing
contributions  to  the  cross section  from   few-nucleon groups.  The
reliability of  the predictions  is verified by   considering together
with $\bar{\mrm{p}}$~production also  the results  for other particles
($\pi^\pm$, $\mrm{K}^\pm$)   which could be  well  described in a wide
energy region.

\ack
One   of  the authors  (HM) would  like  to   thank W~Enghardt for the
promotion of this study.

\appendix
\section*{Appendix}
\setcounter{section}{1}

The $n$ particles of the phase-space (\ref{phase-space1}) of the final
state of the reaction are  divided into two groups, the decay products
of  the   residual  nucleus  and  the  particles   emerging  from  the
projectile-cluster interaction.   In a  first step the  decay products
are  separated  by  introducing   their  invariant  mass  $M_{R}$  and
four-momentum $P_{R}$ via the following identities:
\begin{equation} \label{m:dMR}
  1=\int \di M_{R}^{2}\; \delta (P_{R}^{2}-M_{R}^{2})
\end{equation}
\begin{equation} \label{m:d4PR}
  1=\int \di^{4}P_{R}\; \delta ^{4}(P_{R}-\sum _{i=1}^{n_{R}}p_{i})
\end{equation}
with the definition
\begin{equation} \label{m:PR}
  P_{R}=\sum _{i=1}^{n_{R}}p_{i}.
\end{equation}
Inserting    (\ref{m:dMR})    \(    \dots   \)    (\ref{m:PR})    into
(\ref{phase-space1}) the phase-space factor becomes
\begin{eqnarray}
  \di L_{n}(s) & = & \di M_{R}^{2}\, \di^{4}P_{R}\;
     \delta (P_{R}^{2}-M_{R}^{2})\nonumber \label{dln2} \\
  & & \prod _{i=1}^{n_{R}} \di^{4}p_{i}\, \delta (p_{i}^{2}-m_{i}^{2})\;
     \delta ^{4}(P_{R}-\sum _{i=1}^{n_{R}}p_{i})\nonumber \\
  & & \prod _{i=1}^{n_{P}}\di^{4}p_{i}\, \delta (p_{i}^{2}-m_{i}^{2})\;
     \delta ^{4}(P_{P}-\sum _{i=1}^{n_{P}}p_{i}).\label{M:dLnR0} 
\end{eqnarray}
Taking into account that for on-shell particles the identity
\begin{equation} \label{M:delta}
\int \di^{4}p_{i}\, \delta (p_{i}^{2}-m_{i}^{2})=
   \int \frac{\di^{3}p_{i}}{2e_{i}}
\end{equation}
is valid  and considering  the definition (\ref{phase-space1})  of the
n-particle phase-space equation~(\ref{M:dLnR0}) can be written as
\begin{equation} \label{m:dLnB1}
  \di L_{n}(s)=\di M_{R}^{2}\frac{\di^{3}P_{R}}{2E_{R}}\;
     \di L_{n_{R}}(M_{R}^{2})\; \di L_{n_{P}}(P_{P}^{2}),
\end{equation}
with $n_{P}=n-n_{R}$  being the number of particles  emerging from the
participant interaction. The invariant  mass of this participant group
is   given   by   $M_{P}=\sqrt{P_{P}^{2}}$  with   the   four-momentum
$P_{P}=P-P_R$.  The  separation into  two groups makes  sense, because
the distribution of the momentum $\vec{P}_{R}$ can be derived from the
internal momentum distribution  $\rho (\vec{P}_{R}^{2})$ of the target
nucleus.

Inserting the matrix  element~(\ref{ME}) and the momentum distribution
$\rho (\vec{P}_{R}^{2})$ into  equation~(\ref{m:dLnB1}) and taking the
definitions (\ref{dWR}) and (\ref{dWP})  into account we arrive at the
expression~(\ref{dWaz1}) for  the relative probabilities  of the final
channels $\di W(s;\az)$.

\section*{References}
\bibliographystyle{JourPhysG}
\bibliography{COSY,hadMod,hmbib,hmhelp,nucData,frag,Monte_Carlo}
\end{document}